\documentstyle{article}

\begin{document}
\sloppy

\newcommand{\bbb}{$\spadesuit \spadesuit $}

\newcommand{\aaa}{$\spadesuit \heartsuit \spadesuit $}

\newcommand{\propersubset}{\subset}
\newcommand{\la}{\lambda}
\newcommand{\f}{\varphi}
\newcommand{\de}{\Delta}
\newcommand{\al}{\alpha}
\newcommand{\be}{\beta}
\newcommand{\e}{\in}
\newcommand{\th}{\theta}
\newcommand{\spq}{\stackrel{o}\rightarrow}
\newcommand{\s}{\rightarrow}
\newcommand{\N}{{\bf N}}
\newcommand{\R}{{\bf R}}
\newcommand{\PR}{{\bf PR}}
\newcommand{\FR}{{\bf FR}}
\newcommand{\FPRED}{{\bf FPRED}}
\newcommand{\RPRED}{{\bf RPRED}}
\newcommand{\UU}{{\cal U}}
\newcommand{\OO}{{\cal O}}
\newcommand{\x}{{\bf x}}
\newcommand{\SP}{\;}
\newcommand{\y}{{\bf y}}
\newcommand{\rem}{{\bf !!!! REMINDER }\\}
\newcommand{\COM}[2]{{#1 \choose #2}}
\newcommand{\ie}{\mbox{i.e.}}
\newcommand{\io}{\mbox{i.o.}}
\newcommand{\sstar}{{\Sigma^*}}
\newcommand{\sinf}{{\Sigma^\infty}}

\def\RR{I\!\!R}
\def\NN{I\!\!N}
\def\CC{I\!\!\!\!C}
\def\QQ{I\!\!\!\!Q}
\def\ZZ{Z\!\!\!Z}
\def\Ha{I\!\!H}
\def\FF{I\!\!F}
\def\Po{I\!\!P}
\def\TT{T\!\!\!T}
\def\L{{\cal L}}

\begin{flushright}
\it {Those who most ignore, least escape.}\\
David Hawkins
\end{flushright}

\bigskip

{\let\newpage\relax
\title{{\bf Strong Determinism vs.  Computability}\thanks{This work has
been partly done while
the first author has visited Bucharest University and
the University of Technology Vienna, and the fourth author has
visited the University of  Auckland. The work of the
        first and fourth authors has been
         supported, in part, by Auckland University Research Grants
         A18/XXXXX/62090/3414012, A18/XXXXX/62090/F3414030.}
          }
\author{Cristian Calude,\thanks{Computer Science Department, The University of
Auckland, Private Bag 92109, Auckland, New Zealand; email:
cristian@cs.auckland.ac.nz.}
\hspace*{2mm}
Douglas I. Campbell,\thanks{Computer Science and Psychology
 Departments, The University of
Auckland, Private Bag 92109, Auckland, New Zealand; email:
dcam03@cs.auckland.ac.nz.}
\hspace*{2mm}
 Karl Svozil,\thanks{Institut f\"ur Theoretische Physik,
   University of Technology Vienna,  Wiedner Hauptstra\ss e 8-10/136,
 A-1040 Vienna, Austria; email:  svozil@tph.tuwien.ac.at.}
 \hspace*{2mm}
Doru \c{S}tef\u anescu\thanks{Department of Mathematics, Faculty
of Physics, Bucharest University, P.O.Box 3995, Bucharest 39, Romania;
email:  stef@imar.ro.}
}


\maketitle}

\thispagestyle{empty}
\begin{abstract}
Are minds subject to laws of physics? Are
 the laws of physics computable? Are conscious thought processes computable?
Currently there is little agreement as to what are the right answers to these
questions.
Penrose (\cite{penrose0}, p. 644) goes one step further and asserts that:
{\it  a radical new theory is indeed needed, and I
am suggesting, moreover, that this theory, when it is found, will be
of an essentially non-computational character.}
The aim of this paper is three fold: 1) to examine the incompatibility between
 the hypothesis of strong determinism and computability,  2) to give
new examples of uncomputable physical laws, and 3) to discuss the relevance of
G\" odel's Incompleteness Theorem in  refuting  the claim
that an algorithmic theory---like strong AI---can provide an adequate
theory of mind. Finally, we question the adequacy of the theory
of computation to   discuss physical laws and thought processes.
\end{abstract}

\section{Introduction}
Penrose \cite{penrose} (see also \cite{penrose0}) has discussed a new point
of view concerning
the nature of physics that might underline conscious thought processes.
He has argued that it might be the case that some physical
laws are
not computable, i.e. they cannot be properly simulated by computer; such
laws can
most probably arise on the ``no-man's-land" between classical and quantum
physics.
Furthermore, conscious thinking is a non-algorithmic activity. He is
opposing both
 strong AI  (according to which the brain's action, and, consequently,
conscious perceptions and intelligence, are manifestations of computer
computations,
Minsky \cite{minsky,minsky1}), and Searle's   \cite{searle}
contrary viewpoint (although  computation does
not  in itself evoke consciousness,   a computer might nevertheless
simulate the action of a brain
 mainly due to the fact that the human brain is a
physical system
behaving according to (computable)  mathematical ``laws").

The aim of this paper is to examine the incompatibility between
the hypothesis of strong determinism and computability,  to give
new examples of uncomputable physical laws, and  to discuss the relevance of
G\" odel's Incompleteness Theorem in  refuting  the claim
that an algorithmic theory---like strong AI---can provide an adequate
theory of mind.
Our starting point is the following paragraph  from Penrose \cite{penrose}
p.560:
\begin{quote}
It seems to me that if one has strong determinism, but {\it without} many
worlds,
then the mathematical scheme which governs the structure  of the universe would
probably {\it have} to be non-algorithmic. For otherwise one could in
principle calculate what one was going to do next, and then one could `decide'
to do something different, which would be an effective contradiction between
`free will' and the strong determinism of the theory. By introducing
non-computability
into the theory one can evade this contradiction---though I have to confess
that I
 feel somewhat uneasy about this type of resolution, and I anticipate
something more
subtle for the {\it actual} (non-algorithmic!) rules that govern the way
that the world works!
\end{quote}

\section{From Boscovich to G\"odel}
Perfect determinism was considered earlier by Boscovich \cite{boscovich},
Leibniz and Laplace
(see Barrow \cite{barrow}). The main argument is similar to the one used by
Penrose:
if all our laws, say, of motion, were in the form of equations which
determine the future
uniquely and completely from the present, then a ``superbeing"  having a
perfect knowledge
of the starting state would be able  to predict the entire future. The
puzzling consequence
appears as soon as one tries  to carry out this prediction!

G\" odel was interested in this problem as well.
According to notes taken by Rucker   (\cite{rucker1}, p.181)  G\" odel's
point of view
is the following:

\begin{quote}
It should be possible to form a complete theory of human behaviour, i.e. to
predict from the
hereditary and environmental givens what a person will do. However, if a
mischievous
person learns of this theory, he can act in a way so as to negate it. Hence I
conclude that such a theory exists, but that no mischievous person will
learn it. In the same
way, time-travel is possible, but no person will ever manage to kill his
past self.
\end{quote}

And he continues:

\begin{quote}
There is no contradiction between free will and knowing in advance
precisely what
one will do. If one knows oneself
completely then this {\it is} the situation. One does not deliberately do the
opposite of what one wants.
\end{quote}

\section{Strong Determinism}

According to Penrose (\cite{penrose}, p. 558-559)  strong determinism
\begin{quote}
is not just a matter of the future being
determined by the past; the {\it entire history of the universe
is fixed}, according to some precise mathematical scheme, {\it for all
time}.
\end{quote}

Thus strong determinism  is a variant of
Laplace's scenario,\footnote{{\em  ``A thing cannot occur without a cause
which produces it"}.}   according to which the stage is set at the
beginning and everything follows ``mechanistically'' without the
intervention of God, without the occurrence of ``miracles''
 (cf.
Frank
\cite{frank}).

Strong determinism
does not imply a computable Universe,
 as it says nothing about the computability of
initial conditions  or of physical laws.\footnote{Assuming
the Church-Turing Thesis, this is
equivalent to saying  that the laws of nature correspond to recursive
functions.}

 Let us discuss this in the context of the
computer science. Any program $p$ requiring some particular input $s$
can be rewritten into a new program $p'$ requiring no (the empty list
$\emptyset $) input. This can for instance been realized by coding the
input $s$ of $p$ as constants of $p'$.
Likewise, any part of $p'$ can be externalized as a subprogram $s$,
whose code can then be identified with an input for the new program $p$.
In this sense, the terms effective computation and initial value are
interchangeable and the naming merely a matter of convention.
Therefore, if strong determinism leaves unspecified the computability of
initial values serving as input for recursive natural laws, it may as
well leave unspecified the recursion theoretic status of  natural
laws.

All this sounds rather abstract and mathematical, but the emergence of
chaotic physical
motion has confronted the physics community with the theoretical question of
 whether or not to accept the classical
(i.e., non-constructivist) continuum.
As envisioned by  Shaw \cite{shaw} and Ford \cite{ford}, along with
many others, ``classical chaos'' emerges by the effectively
computable ``visualization'' of the incompressible algorithmic
information of the initial values. Thereby, the classical continuum
serves as an ``urn'' containing (almost, i.e., with probability one)
only (uncomputable)  Martin-L\"of/Chaitin/Solovay random elements.
With probability one, the physical system ``chooses'' one random
element of the continuum
``urn'' as its initial value.
In this sense, chaotic dynamics expresses almost a tautology:
put Martin-L\"of/Chaitin/Solovay
 randomness in, get chaotic motion out.
The non-tautologic feature is the ``choice'' of one
element of the classical (i.e., non-constructivist) continuum.
In order to be able to choose from non-denumerable many uncomputable
objects, the axiom of choice has to be assumed. But then, one is
confronted with ``paradoxical'' constructions utilizing this axiom
(cf.  Wagon
\cite{wagon,svozil}).
 In particular, one could transform every given physical
object into any other physical object (or class of objects) in three
processing steps:
\begin{description}
\item[$\bullet$]
decompose the original object into a finite number of pieces;
\item[$\bullet$]
apply isometric transformations such as rotations
and translations to the pieces; and finally,
\item[$\bullet$]
rearrange them into the final form.
\end{description}
This might  be the ultimate production belt: one can obtain an arbitrary
number of
identical copies from a single prototype!
We mention this utopy here not because of immediate technological
applicability but  to point out the type of shock
to which the physics community is going to be exposed
if it pretends to keep
the
``skeleton in the closet of continuum physics''.
Indeed, all the following examples of strong
determinism clashing with uncomputability and randomness originate in
the assumption of the appropriateness of the classical continuum for
physical modelling.

Quantum theory does not offer any real advancement over classical physics
in this
respect. It is a  ``half-way'' theory, in between the continuum and the
discrete.
 As Einstein put it \cite{ein1},
\begin{quote}
 There are good reasons to assume that nature
cannot be represented by a continuous field.   From quantum theory it could be
inferred with certainty that a finite
system with finite energy can be {\em completely} described by a {\em
finite} number of (quantum) numbers.
This seems not in accordance with continuum theory and has to stipulate
trials to describe reality by purely algebraic means.
Nobody has any idea of how one can find the basis of such a
theory.
\end{quote}

Continuous hidden variable models of quantum mechanics such as Bohm's
model \cite{bohm} operate with pseudo-classical particles. The
real-valued initial position of a Bohmean particle, for instance, is
Martin-L\"of/Chaitin/Solovay
random with probability one. The particles move through
computable quantum potentials.
As in chaos theory, the random occurrence of single particle
detections  originates again in the assumption of the classical
continuum.
{}From this point of view, the Bohmean model of quantum mechanics is not a
``mechanistic'' theory, although its evolution laws might be recursive.

Everett's many-world interpretation of quantum mechanics
\cite{everett} is not much of an advance either. It saves
the strong determinism by abandoning the wave function collapse at
the price of a Universe branching off into (sometimes uncountable) many
Universes at any measurement or beam splitter equivalent.
Currently, there is very little knowledge concerning the computational
status of the
wave function\footnote{See Pour-El and Richards \cite{pour-el},
and the objections in Penrose \cite{penrose}, and Bridges
\cite{bridges1}.} or continuous observables. Implicitly, the
underlying sets are the classical (i.e., non-constructive) continua.

\section{Is Description Possible?}

Can a system contain a description of itself?
Of course, no finite system can contain
itself as a proper part.
What we mean by ``description'' here is an {\em algorithmic
representation} of the system. Such an algorithmic representation could
be interpretable as  a ``natural law''  since it should allow the effective
simulation of the system from within the system.

Von Neumann
\cite{burks}
was concerned with the question of self-description in
the context of the self-reproduction of (universal) automata.
His Cellular Automaton model was inspired by organic life-forms, and the
description ``blueprint'' for self-reproduction was inspired by the DNA.
Today, automaton self-reproduction is just one application
of Kleene's fixed-point theorem
\cite{rogers,odi:89}.

 Von Neumann realized that there must be a difference between an
``active'' and a ``passive'' mode of self-description.
The ``passive'' description is given to the system by some
God-like external agent or oracle. It is then
possible for a finite system to contain such a ``passive''
representation of itself within  itself
as a proper part. Based on this description, the system is capable of
simulating itself.\footnote{Certain prediction tasks cannot
be speeded up, though;
see the
discussion below.}
Such a self-description in general cannot be obtained ``actively'' by
self-inspection.
The reason for this is computational complementarity \cite{moore,svozil}
and the recursive unsolvability of the rule inference problem
\cite{gold,svozil}.

\section{Is Prediction Possible?}
   Is there any  incompatibility between the strong determinism
and computability, as Penrose  suggests?  Is it indeed impossible
for a person to
``learn his own theory" (G\" odel)?

Let us  assume that we have both strong determinism
and
computable
physical laws.  For the remainder of this paper we fix a finite
alphabet $A$ and denote by $A^*$ the set of all strings over $A$; $|x|$ is the
length of the string $x$.
A {\it (Chaitin) computer} $C$ is  a partial recursive function carrying
strings  (on  $A$) into strings such that the domain of
$C$ is prefix-free, i.e. no admissible program can be a prefix of another
admissible program. If $C$ is a  computer, then $T_C$ denotes its
time complexity, i.e. $T_C(x)$ is the running time of $C$ on the entry $x$, if
$x$ is in the domain of $C$; $T_C(x)$ is undefined in the opposite case.
One can prove Chaitin's
Theorem (see,
for instance, Chaitin \cite{ch8,ch9}, Calude \cite{cris}, Svozil \cite{svozil})
stating
the existence of a {\it universal computer} $U$ such that for every
computer $C$ there
exists a constant $sim(U,C)$---which depends upon $U,C$---such that in case
$C(x)=y$,
 there exists\footnote{And can be effectively constructed.} $x'$ such that
\begin{equation}
\label{univ}
U(x')=y,
\end{equation}
 \begin{equation}
\label{optimal1}
  |x'| \leq |x| + sim(U,C).
\end{equation}

 Assume,  now, for the sake of a contradiction, that an
``algorithmic prediction" is possible. Then
the universal computer can simulate the predictor, so it can  itself act as
a predictor. What
does this mean? The computer $U$ can simulate every other computer
(\ref{univ}),
in a shorter time.
Formally, to equation (\ref{univ})  we add
\begin{equation}
\label{optimal2}
  T_U(x') < T_C(x).
\end{equation}

Now, let us examine the possibility that $U$ is a predictor. For every
string $x$ in the domain
of $U$  let
\begin{equation}
\label{time}
t(x)= \min\{ T_U(z)\mid z\in A^\ast , U(z)=U(x)\},
\end{equation}
i.e. $t(x)$
is the minimal running time necessary for $U$ to produce
$U(x)$.\footnote{Actually,
$t(x)$  is not computable.}

Next define
the {\em temporal canonical program (input)}
associated with $x$ to be the first string (in quasi-lexicographical order)
$x^{\#}$  satisfying
the equation (\ref{time}):
\[x^{\#} = \min\{z \in dom(U)\;|\; U(z)=U(x), T_U(z)=t(x)\}.\]
So,
\[U(x^{\#}) = U(x), \mbox{  and  } T_U(x^{\#}) = t(x).\]
As the universal computer $U$  is a  predictor itself, and for itself, it
follows
from
(\ref{optimal2})
that there exists a string $x'$ such that $U(x')=U(x^{\#})=U(x)$, and
$T_U(x') < T_U(x^{\#}) = t(x)$,
which is false. Therefore, every
universal predictor is ``too slow'' for certain tasks, in particular,
predicting ``highly time-efficient'' (or, alternatively, ``highly
time-consuming'') actions of itself.\footnote{For an early
investigation of a  forecast inspired by recursion
theory  see Popper
\cite{popper}.}

The reason for the above phenomenon can be illustrated by
showing the existence of `` small-sized" computers requiring ``very large''
running times.
To this aim we  use
Chaitin's version of the Busy Beaver function
$\Sigma$. Denote
by $H$ Chaitin complexity (or, algorithmic information
content), that is  the function defined on (all) strings by the formula
\[H(x) = \min\{|y| \;|\; y \in A^*, U(y)=x\},\]
i.e. $H(x)$ is the length  of the smallest program for the universal
computer $U$
to calculate $x$.
For every natural
$m$ let us denote by $string(m)$ the $m$th string in quasi-lexicographical
order,
and let  $\Sigma (n)$
 be   the largest natural number whose algorithmic information
content is less than or equal to $n$, i.e.
\[\Sigma(n) = \max\{m \;|\; m \in \NN,  H(string(m)) \leq n\}.\]
  Chaitin
(\cite{ch9},  80-82, 189)  has shown
that
  $\Sigma $
grows larger than any recursive function, i.e. for every recursive
function $f$, there exists a natural number $N$, which depends upon $f$,
such that $\Sigma(n) \geq f(n)$, for all $ n \geq N$: indeed, any program of
length $n$ either halts in time less than $\Sigma(n + O(1))$, or else
it never halts.

As $H(string(\Sigma(n))) \leq n$, it follows that $U(y_n)=string(\Sigma(n))$,
for some string $y_n$ of length less than $n$. This program  $y_n$ takes,
however, a huge amount of time to halt: there is a constant $c$ such that
for large enough $n$, $U(y_n)$ takes between $\Sigma(n-c)$ and $\Sigma(n+c)$
units of time
to halt. To conclude, the equation (\ref{univ})
 is compatible with   (\ref{optimal1})  (Chaitin's Theorem), but
incompatible with (\ref{optimal2}).

Computation is a physical
process, inevitably  bound to physical degrees of freedom; all known
physical laws, in
turn,
are ultimately expressible by algorithms for information processing (i.e.,
they are computable). The above
discussion revealed some mathematical limits; they can be completed with pure
physical limits, as discovered by Mundici \cite{mundici}.\footnote{Gandy
\cite{gandy1,gandy2}
 has put forward  related  arguments  imposing limitations to
mathematical knowledge by the finiteness of physical objects.} Due to the
fact that
 every computer is
subject to the irreversibility and uncertainty of time-energy, and
maximality of the speed light, one can derive the following result: {\it
The total time
$t$ and energy $E$ spent for every computation consisting of $n$ steps satisfy
the inequality:}
\[t \geq n^{2} \frac{h}{2\pi E},\]
{\it where $h$ is Planck constant.} For instance, it follows that
computations involving
more than $10^{30}$
steps are
 infeasible.

This suggests that even inthe case the Universe is deterministic and unique,
and its underlying laws are algorithmic, an algorithmic prediction is
impossible. It justifies
also  G\" odel's
 claim
according to which
``no person will ever learn his theory"
in spite of the fact  that such a theory might exist.

\section{Uncomputability and Randomness: Two Examples}

Various  physical problems  lead to the question whether a
function, in a certain a class, has a real root. Results due to
Richardson \cite{richardson}, Caviness \cite{caviness}, Wang \cite{wang} (see
also Matijasevi\v{c} \cite{matia})
show that for a large class of well-defined functions such a problem
is not algorithmically solvable.  Da Costa and  Doria  \cite{costa}
have proven some undecidability  results in physics using this tool.
A different approach, based on Specker's Theorem, was developed by Pour-El
and Richardson \cite{pour-el}.
In this chapter we shall build on the work of Richardson, Wang,
and Chaitin to show that two problems in
elementary physics are undecidable and display pure randomness.

\subsection{Richardson-Wang and Chaitin Theorems}

An {\it exponential Diophantine equation} is of the form
\[E_1(x_1,\ldots ,x_m) = E_2(x_1,\ldots ,x_m),\]
where $E_1, E_2$ are expressions constructed from variables and natural
numbers,  using
addition, multiplication, and exponentiation. The equations which do not
make use of
exponentiation are called  {\it Diophantine equations}. Fermat's famous
equation
\[(p+1)^{s+3} + (q+1)^{s+3} = (r+1)^{s+3},\]
is an example of an exponential Diophantine equation. For every fixed $s$,
the above
equation is  a Diophantine equation, for instance, the equation
\[(p+1)^{3} + (q+1)^{3} = (r+1)^{3}.\]

By a {\it family} of (exponential) Diophantine equations we understand an
(exponential)
Diophantine equation
\begin{equation}
\label{family}
E_1(a_1, \ldots , a_n,x_1,\ldots ,x_m) = E_2(a_1, \ldots , a_n,x_1,\ldots
,x_m),
\end{equation}
in which the set of all variables $a_1, \ldots , a_n,x_1,\ldots ,x_m$ is
divided into two classes,
{\it unknowns,} $x_1,\ldots ,x_m$, and {\it parameters,} $a_1, \ldots , a_n$.
A set $S \subset \NN^n$ is called  {\it (exponential) Diophantine} if there
exists a
family of (exponential) Diophantine  equations (\ref{family}) such that
\[S = \{(a_1, \ldots , a_n) \in \NN^{n}|
E_1(a_1, \ldots , a_n,x_1,\ldots ,x_m) = E_2(a_1, \ldots , a_n,x_1,\ldots
,x_m),\]
\[\mbox{   for some naturals   }
x_1,\ldots ,x_m\}.\]

 Due to work of  Davis, Matijasevi\v{c}, Putnam, Robinson  (see
Matijasevi\v{c} \cite{matia})
the following classes of sets were shown to coincide:
1) the class of recursively enumerable sets,
2) the class of exponential Diophantine sets,
3) the class of  Diophantine sets.

By virtue of the existence of recursively enumerable sets which are not
recursive (see, for instance,
Calude \cite{ca1}) we deduce that the problem of testing whether an
arbitrary (exponential)
Diophantine equation has a solution (in natural numbers) is recursively
undecidable.\footnote{This
solved in the negative Hilbert's Tenth Problem.}
A universal (exponential) Diophantine set, i.e. a set which ``codes" all
(exponential) Diophantine sets is recursively enumerable, but not recursive.

In contrast with the case of (exponential) Diophantine equations---dealing
with solutions
in natural numbers---the problem of deciding the solvability of polynomial
equations  with integer coefficients in {\it real} unknowns is {\it
decidable}. In the unary
case this can be done by the well-known Sturm method; in the  general case
one have to use
Tarski's method \cite{tarski}. To get undecidability  we have to allow the use
of some other functions; an easy way to achieve this is to consider the
addition, multiplication,
composition and the sine  function, all rationals and $\pi$.

For our aim it is convenient to reformulate Richardson \cite{richardson}
and Wang \cite{wang} results as follows.  We define, for every natural $n
\geq 1$, $\Delta_n$
to be
the minimal (with respect to
set-theoretical inclusion) family of expressions
 which contains
all rationalsand $\pi$,
the variables $x_1, \ldots ,x_n$,
the functions $sin(x)$ and $e^x$,
and which is closed under the operations of addition, multiplication, and
composition.

The following predicates are recursively undecidable:
\begin{itemize}
\item
For every $G(x_1) \in \Delta_1$, ``there exists a real number $r$ such that
$G(r) = 0$".
\item
For   every $G(x_1) \in \Delta_1$, the predicate ``the integral
 $\int_{- \infty}^{+\infty} [(x^2 + 1) G^{2}(x)]^{-1} dx$
is convergent".
\end{itemize}

Following Chaitin \cite{ch8,ch9} we do not ask whether an arbitrary
Diophantine equation has a solution, but rather whether it has an infinity of
solutions. Of course, the new question is still undecidable. In the former
case  the answers to such questions are  not independent\footnote{The
reason is simple:
we can determine which equations have a solution if we know how many of
them are solvable.},
but in the later one the answers can be independent in case the equation is
constructed
properly. Actually Chaitin has effectively constructed such an exponential
Diophantine equation (see his last Lisp construction in \cite{ch10})
with the property that the number of solutions jumps from finite to
infinite at {\it
random} as a certain fixed parameter is varied. Actually, saying that the
``number of solutions jumps from finite to infinite at {\it
random}" is not a figure of speech, it is just a remarkable technical
statement:
if  the parameter $n$ takes the values $ 1,2, \ldots$, and $\omega_n = 0$
in case the corresponding
equation has finitely many solutions, and $\omega_n= 1$, in the opposite
case, then
the sequence $\omega_1\omega_2 \cdots \omega_i \cdots$ is random in
Martin-L\" of/Chaitin/Solovay
sense; see Calude \cite{cris}.
 The real number number
\[\Omega = 0.\omega_1\omega_2 \cdots \omega_i \cdots\]
represents the halting probability of a universal  computer. In case we
assume the
hypothesis of strong determinism, $\Omega$ has also a ``physical"
significance: it represents
a constant of the Universe.\footnote{There is something attractive
about permanence.} The number $\Omega$ is not invariant under changes of
the underlying universal  computer.
However, all ``constants" $\Omega$ share a number of fascinating properties
(see, for instance,
Calude \cite{cris}); these changes  might be similar to changes
of other ``constants of Nature", as Newton's gravitational constant, the
charge of an electron or the
fine-structure constant, under certain circumstances (changing the number
of dimensions of the space, for instance).

\subsection{One-dimensional Heat Equation}

Improper integrals, for example,
Fourier and Laplace transforms, play a particularly important role
in modelling physical phenomena (see, Courant,  Hilbert
\cite{ch}, \c{S}tef\u anescu \cite{ds}).
Two examples involving the Laplace transform  illustrate
uncomputability and randomness.

Let us first consider the heat conduction on an infinite slab. It is described
by the one-dimensional heat equation:

        \begin{equation}\label{one-heat}
                \left\{\begin{array}{lc}
\displaystyle\frac{\partial u}{\partial t} -
\frac{\partial^2 u}{\partial x^2} = 0,  x\in \RR, \; t>0, \\
\\
u(x,0) = f(x), \\
\\
u(x,t) \mbox{ \rm is bounded.}
                \end{array}\right.
        \end{equation}

If $\displaystyle\frac{\partial u}{\partial t}$ and
$\displaystyle\frac{\partial^2 u}{\partial x^2}$ are supposed to be continuous
and bounded, then the solution of  (\ref{one-heat}) may be obtained
via the Laplace transform (see, Friedrichs \cite{kof}):\footnote{Notice that
the solution of the problem (\ref{one-heat}) may be also obtained by means
of the Fourier transform.
It is possible that for some functions $f$ the Laplace (or Fourier)
transform does not exists, and still  (\ref{sol-heat})
verifies  (\ref{one-heat}). }

        \begin{equation}\label{sol-heat}
u(x,t) = \displaystyle\frac{1}{2\sqrt{\pi t}}\int_{-\infty}^{\infty}
e^{-\frac{(x-y)^2}{4t}}f(y)dy.
        \end{equation}

        \subsection{A Problem of Electrostatics}

Let us consider the plane electrostatic problem\footnote{A problem of
electrostatics is plane if there is a distinguished
direction such that all data are constant in this direction and the field
to be determined is also constant in this direction;  Friedrichs
\cite{kof}.}
 on $\RR\times \RR_+$ which
satisfies the boundary potential condition
        \[ \Phi (x,0) = f(x). \]
If $\Phi$ is an electrostatic potential, then the electric field
${\bf E}$ is given by
        \[{\bf E} = - \mbox{\rm grad  } \Phi .\]

If ${\bf D}$ is a plane domain (i.e. an infinitely long cylinder with cross
section ${\bf D}$) bounded by a
surface
 ${\bf C}$ composed of several
conductors\footnote{The conductors are materials which do not exert any
force on charged particles in their interior, but they do so at the
boundary. In a state of equilibrium the charges contained in a conductor
are distributed over the boundary.} at different potentials, then $\Phi$ is
 is a solution of the
system\footnote{The
same system can be derived from conduction of electricity on a conducting
sheet covering the domain ${\bf D}$.}

        \begin{equation}\label{electrost}
                \left\{\begin{array}{lc}
\displaystyle\frac{\partial^2 \Phi }{\partial x^2} +
\frac{\partial^2 \Phi }{\partial y^2} = 0,   (x,y)\in {\bf D},\\
\\
\Phi (x,0) = f(x).
                \end{array}\right.
        \end{equation}

The problem (\ref{electrost}) can be solved via the formalism
of differential forms.\footnote{The
local existence of a potential $\Phi$ is described by the equality
${\bf E} = -{\rm d  }  \Phi$; see Bamberg and Sternberg \cite{bs}.}
The solution of (\ref{electrost}) is given by
        \begin{equation}\label{sol-electrost}
\displaystyle   \Phi (x,y) =
\frac{y}{\pi}\int_{-\infty}^{\infty}\frac{f(t)}{(t-x)^2+y^2} dt.
        \end{equation}

        First we look at the solution of the one-dimensional heat equation
(\ref{sol-heat}).
If $f(y) = (y^2+1)^{-1},$ then, for every fixed $(x_0,t_0)$, the
solution \[u(x_0,t_0) = \frac{1}{2\sqrt{\pi t_0}}\int_{-\infty}^{\infty}
\frac{e^{-\frac{(x_0-y)^2}{4t_0}}}{y^2+1}dy \] is finite.

      Consider now the function $f(y) = e^{y^2}$.
Let $t_0>1$ and $x_0\in \RR$ be fixed. Then
\[ e^{-\frac{(x_0-y)^2}{4t_0}} f(y) > e^{y^2 - \frac{(x_0-y)^2}{4}} =
e^{\frac{3}{4}y^2 + \frac{x_0y}{2} - \frac{x_0^2}{4}}.\]
       For  fixed $x_0$,
$\lim_{y\rightarrow\infty}\frac{3}{4}y^2 + \frac{x_0y}{2} - \frac{x_0^2}{4}
= \infty,$
so the integral
\[ \int_{-\infty}^{\infty} e^{-\frac{(x_0-y)^2}{4t_0}} f(y) dy\]
is divergent.

        If  $f(y) = (y^2+1)^{-1}H^{-2}(y)$  then, for every fixed
$(x_0,t_0)$, we get the solution
         \[ u(x_0,t_0) = \frac{1}{2\sqrt{\pi t_0}}\int_{-\infty}^{\infty}
\frac{e^{-\frac{(x_0-y)^2}{4t_0}}}{(y^2+1)H^2(y)}dy =\\
 \frac{1}{2\sqrt{\pi t_0}}\int_{-\infty}^{\infty}
\frac{1}{(y^2+1)K^2(y)}dy.\]
In case $H$ was in $\Delta_1$, then $  K$ is in $\Delta_1$ as well. So, the
problem
to test, for fixed $(x_0,t_0)$,   whether
 the solution $u(x_0,t_0)$ is finite or not for an arbitrary function
$H\in\Delta_1$,
is recursively undecidable.

Using Chaitin's construction we can exhibit a
sequence of functions  $H_i\in\Delta_1$ such that the induced sequence
$c_1c_2\cdots c_i\cdots$, $c_i = 0$, if the corresponding solution is finite,
$c_i = 1$,  in the opposite case, is random. So, in the space of all solutions
of (\ref{sol-heat}) there are areas in which convergence and divergence
alternate
in a pure random way.

        Similar results can be obtained for the solution of the electrostatic
plane problem.
For fixed $x_0,y_0$, $y_0\ne 0$, the solution
(\ref{sol-electrost}) can  be represented as
\begin{equation}\label{undec-sol}
        \displaystyle \Phi(x_0,y_0) = \frac{1}{\pi
y_0^2}\int_{-\infty}^{\infty}
\frac{f(y_0u+x_0)}{u^2+1} du.       \end{equation}
        If $f(x) = G(x)^{-2},$
where $ G$ is a function in $\Delta_1$,
then  the the problem of testing whether $\Phi(x_0,y_0)$  is finite or not
is recursively undecidable. Again,  we can effectively construct
a sequence of solutions displaying pure randomness, i.e. for which
the sequence of answers to the convergence problem is random.

\section{Incompleteness}

In a remarkable  paper entitled {\it Intelligent
Machines}\footnote{This paper has attracted less interest than {\it
Computing Machinery and
Intelligence} (\cite{tu2}, 133-160); for instance, Penrose does not quote
it at all.}
 (\cite{tu2}, 107-127)
Turing investigates the possibility as to whether machines, i.e. computers,
might
show intelligent
behaviour.
He considers the argument that machines are inherently incapable of exhibiting
human-like intelligent behaviour, because human mathematicians are capable
of determining the truth or falsity of mathematical statements in a way
that machines, as embodiments of formal systems that are subject to the
limitations of G\" odel's Incompleteness Theorem, cannot.
Turing notes that G\" odel's Incompleteness Theorem
\begin{quote}
 rests  essentially on the condition
that   the machine  must not make  mistakes.
But this is not a requirement
for intelligence.
\end{quote}

He is suggesting that machines might perhaps equal human mathematicians if
they were equipped
with a human-like capacity to make mistakes.

The   analysis of predictability outlined in this paper
 is subject to Turing's objection regarding mistakes. Accordingly,  we
address the
following  question:
Is Turing's argument irrefutable?

 At a first sight,  requiring the absence of  mistakes
might seem to be  overly  restrictive.
But how can a mistake-making machine  be constructed?
 Where should we place the border between
``admissible"
and ``non-admissible" mistakes in order to preserve  the ``intelligibility"
of our Universe.
How can a mistake-making machine
discover the regularities, common factors, recurrences, and  implications,
which tell us
what  things are and how are they going to be in the future?
According to Barrow
(\cite{barrow} p. 269):

\begin{quote}
 the intelligibility of the world amounts to the fact that we find it to be
algorithmically compressible.  We can replace sequences of facts and
observational data by
abbreviated statements which contain the same information content. These
abbreviations we often call ``laws of Nature".
\end{quote}

However, we know that a total compression of the Universe is not
actually possible as the
existence of chaotic processes points out (Chaitin \cite{ch8,ch9},
Rucker \cite{rucker1},  Svozil \cite{svozil,svozil1,svozil2},
Calude \cite{cris}, Calude and Salomaa \cite{caludesalomaa}). How can we
describe  seemingly random processes in nature and reconcile them with
supposed order? How much can a given piece of information be compressed?
 Calude and Salomaa \cite{caludesalomaa} have suggested that the Universe
is actually globally
{\it random}, and, consequently, locally  {\it ordered}. The Universe, like
any network-like structure can
be seen both at  local  and  global levels. Local properties require only a
very nearsighted
observer---and for this level, science is indeed very useful and
successful---but global properties
are much  more difficult to ``see", they
need a sweeping vision. For instance, the overall shape of a
spiderweb is a global property, while the average  number of lines meeting
a vertex is
a local characteristic.

The relevance of G\" odel's Incompleteness Theorem \cite{goedel}
 argument has been
questioned by different authors, especially by
Boolos, Chalmers, Davis and Perlis (see \cite{penrose0}; it contains also
Penrose's reply).
In our opinion, Turing's critique---mentioned above---is the most
substantial. It
questions the status of G\" odel's  famous unprovable statement: is this
unprovable
statement---seen to be ``true" by Penrose---esoteric, accidental? Does
the incompleteness  phenomenon have any relevance
for a scientist's daily life?  This is a rather delicate
question. If we adopt a topological  point of view (see Calude, J\"
urgensen, Zimand \cite{cjz}), then
incompleteness is a rather common, pervasive phenomenon: the set of true,
but unprovable
statements is topologically ``very large", i.e. with respect
 to any reasonable topology
the set of true and unprovable statements of a sufficiently rich,
sound, and recursively axiomatizable theory
is dense and in many cases even co-rare. It is important to notice that
the above result holds true not only globally, but even for
``fixed" problems. For instance, the  halting problem: there exists
a large set of  true, but unprovable, statements stating that some Turing
machine
will never halt on a fixed entry.

The natural way to model ``admissible mistakes" is to work with
probabilistic Turing
machines\footnote{This type of machine is sometimes called a {\it Monte
Carlo} algorithm.}
instead of (ordinary) Turing machines. A probabilistic Turing machine has
some distinguished states acting as ``coin-tossing states" for which
the finite control   specifies $p \geq 2$ possible next states. The
computation is
deterministic except that in the distinguished states the machine uses the
output of a random
experiment to decide among the $p$ possible next states. So, a
probabilistic Turing machine
can make mistakes; the output is not ``truly correct", but ``correct within
a probability".
Classical results
due to   De Leuuw,  Moore,  Shannon, and Shapiro \cite{dmss} and Gill
\cite{gill}
show that   the class of functions computed by
probabilistic algorithms  coincides with the class of
recursive functions. The difference is only in complexity: if we do not
insist on a
{\it guarantee}, then sometimes it is possible to compute faster. All
results pertaining
incompleteness,   previously discussed, remain valid, so  it appears that
Turing's
objection cannot be supported anymore: this probabilistic space inherits the
non-computability of the deterministic one.

\section{Computability}

Is  the theory of computability (recursion theory)\footnote{A truly
remarkable achievement of
modern mathematics is
the discovery of recursive
(or, computable) functions, i.e. functions which can be computed by
algorithms.
Within the realm of this
theory it is possible to prove the existence of functions that are not
computable
by any algorithm whatsoever.
The theory of computability  has not yet become part
of mainstream physics, but it can serve perfectly well  as a guiding
principle to hitherto informal notions such as ``determinism''.
}
 an appropriate framework to discuss physical laws and thought processes?
It is not unreasonable to suspect that the notion of
computation will play a major role in future research in the
natural sciences; however, the global picture is more complex than it
appears on a first analysis.

Recursion theory is useful for proving the existence of uncomputable
physical laws.
If we are interested in ``useful" physical laws, i.e. laws
which can be effectively used for practical purposes, then the theory of
computation
might not be the appropriate tool. Indeed, it
 may happen that some function is computable, but it is very difficult
to compute,\footnote{Actually, for every computational  measure, for
instance,  time
or space, there exist arbitrarily difficult to compute functions; see
Calude \cite{ca1}.}
or even worse,  that the computable function
 {\it is impossible} to compute  at all. For instance, consider the Continuum
Hypothesis\footnote{There is no
cardinal
number  strictly in between aleph-null, the cardinal of the the set of
natural numbers, and aleph-one, the cardinal of the set of reals.} and the
following function
 \[f(n) = \left\{ \begin{array}{ll}
1, & \mbox{ if the Continuum Hypothesis is true,}\\
0, &  \mbox{ if the  Continuum Hypothesis is false, }
\end{array}
\right.\]
suggested in Bridges \cite{bridges}.
According to classical logic,  $f$ is computable because there exists an
algorithm
that computes it,
i.e. the algorithm that returns either one or zero,  for all non-negative
integers. Deep work
due to G\"{o}del
\cite{goedel3} and Cohen\cite{cohen} shows that neither the  Continuum
Hypothesis nor its negation can be proven within
Zermelo-Fraenkel set
theory augmented with the Axiom of Choice, the standard framework of
classical mathematics,
so we will never know which of the two algorithms---``print one", or
``print zero"---is the right one. We
conclude
that the standard theory of computable functions does not match
computational practice!
The paradoxical nature of this example comes from the underlying logic of
computability.
To handle this problem we have to distinguish between {\it existence in
principle}
and {\it existence in practice}. A possible approach is to consider
provable computable
functions introduced by Fischer \cite{fischer}. A  computable function is
called {\it provable}
with respect to some formal system $S$ which contains second order
arithmetic if there
exists an algorithm which computes it and which can be proven to be total
in $S$. These
functions are interesting because they are  functions we usually work
with in practice, e.g.
in numerical analysis. What do we lose sacrificing all computable functions
in favour of
provable computable ones? Gordon \cite{gordon} has proven  that
this class of functions is a complexity class, i.e. it can be computed with
limited
resources, say in time. Now, if we apply some results in Calude \cite{ca1}
we arrive at the conclusion that there is an essential
 difference between computable functions and provable computable
functions:
in a constructive sense, the former class is of second Baire category (i.e.
large)
 while the later one is meagre (i.e. small). Informally this means that
most computable
functions are not provable computable; the difference between
functions ``computable in principle" and provable computable functions
is significant.\footnote{In this context
it is interesting to  note a result---obtained in 1964---which can be
considered as
 ``Chaitin (very first) Incompleteness Theorem":
{\it For any formal system there is a computable
total function that goes to infinity more quickly than any provably
computable total function in the formal system. }   For the construction we
take
$ F(n)$  to be $ n $   times the maximum of the values of the first $n$
provably
computable total functions for all arguments up to $n$;   "first" means
first in a recursive enumeration of all theorems in the formal system.
This note was has kindly communicated to us in \cite{email}.}

\section{Conclusions}
 The paradox mentioned by Penrose is not real, because ``real predictors"
do not exist.\footnote{Penrose himself seems to have anticipated this.}
This is because
every  (universal) predictor is ``too slow" for certain tasks, in particular
for predicting actions of itself. Two more examples of uncomputability
of physical laws are discussed. Turing's objection concerning
G\" odel's Incompleteness Theorem  is confronted with the
fact that,  from a topological point of view,
the incompleteness
phenomenon is common and pervasive; this result is still true
for probabilistic Turing machines, i.e. for machines allowed
to make ``reasonable" mistakes.
Although we have refuted Penrose's argument that strong determinism
and computability are logically incompatible, we have found independent
reasons to support his conclusion concerning the non-computability of physical
laws.
Finally we are lead to the following question:
is the  theory of computation
 an appropriate framework to discuss physical laws and thought processes?
We argue that for proving non-computability results the answer is affirmative;
for more practical purposes,  in  which we are interested not only in
discovering physical laws, but in using them to make predictions, the answer
might be negative.  Other aspects of the problem, e.g., the role of the
observer
and ``approximation" in making predictions, will be treated in another
paper.

\end{document}